\documentclass{PoS}
\usepackage{amsmath,amssymb,graphicx}

\title{Simulated random surfaces and effective string models in 3D Z(2) gauge theory}

\ShortTitle{Simulated random surfaces and effective string models in 3d Z(2) gauge theory}

\author{Tomasz Korzec\\
        Humboldt Universitat zu Berlin, Institut fur Physik, Newtonstrasse 15, D-12489 Berlin, Germany\\
        E-mail: \email{korzec@physik.hu-berlin.de}}

\author{\speaker{Ulli Wolff}\\
        Humboldt Universitat zu Berlin, Institut fur Physik, Newtonstrasse 15, D-12489 Berlin, Germany\\
        E-mail: \email{uwolff@physik.hu-berlin.de}}

\abstract{All-order strong coupling simulations have been used to derive precise energies of
string states in the confined phase of three dimensional Z(2) lattice gauge theory.
The behavior of the ground state energy is here compared with predictions of effective string theory.
Our new data reported here 
are consistent with known universal terms of the long string length ($L_0$) expansion known
from effective string models in the continuum limit. Our precision is however still not sufficient
to disentangle non-univeral terms of order $L_0^{-7}$.

\begin{flushright} HU-EP-13/43 \end{flushright}
\begin{flushright} SFB/CCP-13-62 \end{flushright}
}

\FullConference{31st International Symposium on Lattice Field Theory LATTICE 2013\\
                 July 29 - August 3, 2013\\
                 Mainz, Germany}

\newcommand{\Omicron}{\mathrm{O}}
\newcommand{\mathe}{\mathrm{e}}
\newcommand{\tmop}[1]{\ensuremath{\operatorname{#1}}}
\newcommand{\tmtextbf}[1]{{\bfseries{#1}}}

\newcommand{\tmtexttt}[1]{{\ttfamily{#1}}}

\begin{document}

\section{Introduction}

Starting from the work of Prokof'ev and Svistunov {\cite{prokofev2001wacci}} a
number of scalar lattice field theories have been simulated by so-called worm
algorithms. An overview and more references can be found in
{\cite{Wolff:2010zu}}. The main ingredient of the method is the replacement of
the original path integral by an untruncated strong coupling series for the
partition function. Contributions are represented by an ensemble of positively
weighted graphs. For any finite lattice these series are convergent for all
values of the bare parameters and may thus be considered as equivalent to the
original systems. To realize the convergence for interesting areas in
parameter space, high orders are important and the summation is only possible
by stochastic methods like Monte Carlo simulation. The observation in
{\cite{prokofev2001wacci}} has been that in such simulations it is highly
profitable to enlarge the graph space from the graphs of the partition
function to those required for the fundamental two point correlation between
arbitrary pairs of sites $u, v$, the `head' and `tail' of the worm. In the
graphical language this introduces a pair of point defects where lines can end
that otherwise have to form closed loops. Although the relevant graphs are
essentially sampled by local deformations only, very little critical slowing
down hampers the computation of relevant observables. Moreover, by sampling
the enlarged graph space, a very efficient computation of the two point
function and quantities like the mass gap becomes possible. As an additional
bonus it is shown in {\cite{Wolff:2008km}} that the exponential decay of such
correlations can be numerically followed to large separations without the
usual degradation of the signal to noise ratio. This is achieved by
introducing suitable weight factors into the graph ensemble that is simulated.
A further advantage of worm methods is that in some cases, in particular with
nonzero chemical potential, a sign problem present in the original formulation
is absent for the equivalent graphs and is hence cured.

The success story outlined above has so far mainly been restricted to
nonlinear sigma models of the O($N$) and CP($\left. N - 1 \right)$ type as
prototype scalar field theories. Even setting aside fermions a natural next
candidate to study are gauge theories. For Abelian theories (Z($N$), U(1)) the
generalization to the strong coupling representation of the partition function
is at least superficially straight forward. Admissible defects are now
superpositions of arbitrary closed loops instead of point defects in the
scalar models. The loop defects are related to Wilson loop observables in a
similar way as the point pairs connect to scalar correlations. A key
difference is now that possible loop defects of arbitrary shape form a very
much larger set than the possible locations of pairs of points. In our recent
article {\cite{Korzec:2012fa}} we have studied Z(2) gauge theory{\footnote{For
earlier related attempts in U(1) gauge theory there are more references in
{\cite{Korzec:2012fa}}, in particular {\cite{Korzec:2010sh}}.}} in three
dimensions, which is dual to the Ising spin (or O(1) sigma) model. In a series
of simulations at the critical point we have experimented with various subsets
of such line defects but have failed to identify a set that is efficient with
regard to slowing down. We have then changed our direction: we have accepted
critical exponents around two but tailored the simulated ensemble to optimize
the estimation of Polyakov line two point functions. Here we have indeed
succeeded in achieving a separation independent signal to noise ratio which
allows for the very precise calculation of the ground state `quark-antiquark'
potential and the properties of string states in a finite volume. In the rest
of this contribution we discuss these results and their matching with
effective string theory predictions. Further details on the algorithmic
aspects of our investigation are summarized, including graphical
illustrations, in the contribution by Tomasz Korzec {\cite{KorzecLat13}}.

\section{Polyakov line correlation function}

Our calculation takes place in Z(2) gauge theory on a periodic lattice of size
$L_0 \times L \times L$. Employing the standard Wilson action defined on
plaquettes the partition function is given by
\begin{equation}
  Z = \sum_{\left\{ \sigma_{\mu} \left( x \right) = \pm 1 \right\}}
  \mathe^{\beta \sum_{x, \mu < \nu} \sigma_{\mu \nu} \left( x \right)}
\end{equation}
where $\sigma_{\mu \nu} \left( x \right)$ is the product of the links
$\sigma_{\mu} \left( x \right)$ around a plaquette. The Polyakov line
correlation function is defined by
\begin{equation}
  G \left( \vec{x} \right) = \langle \pi \left( \vec{x} \right) \pi \left(
  \vec{0} \right) \rangle, \hspace{1em} \pi \left( \vec{x} \right) =
  \prod_{x_0 = 0}^{L_0 - 1} \sigma_0 \left( x_0, \vec{x} \right) .
\end{equation}
In our random surface simulation there are straight line defects of length
$L_0$ at two dimensional locations $\vec{u}$ and $\vec{v}$ which randomly
wander over the lattice {\cite{Korzec:2012fa}}, {\cite{KorzecLat13}}. 
The random surface ensemble is
constructed such that we can prove the relation
\begin{equation}
  G \left( \vec{x} \right) = \rho^{} \left( \vec{x} \right) \frac{\langle
  \langle \delta_{\vec{x}, \vec{u} - \vec{v}} \rangle \rangle}{\langle \langle
  \delta_{\vec{u}, \vec{v}} \rangle \rangle} .
\end{equation}
The function $\rho^{} \left( \vec{x} \right)$ has been incorporated to bias
our updates and is exactly canceled here to yield $G \left( \vec{x} \right)$.
It has been modeled to capture much of the variation of $G$ with $\left|
\vec{x} \left| \right. \right.$ and this is what leads to our favorable signal
to noise behavior, because the Monte Carlo only computes the slowly varying
correction factor that turns $\rho$ into the exact correlation.

On the theoretical side $G \left( \vec{x} \right)$ can be viewed as the
partition functions of the Z (2)  gauge theory with static charges separated
by $\vec{x}$. If we employ the transfer matrix with respect to the temporal
direction we arrive at the formula {\cite{Luscher:2004ib}}
\begin{equation}
  G \left( \vec{x} \right) = \sum_{n \geqslant 0} w_n \mathe^{- V_n \left(
  \vec{x} \right) L_0} \label{pots}
\end{equation}
where $w_n$ are integer multiplicity factors. Assuming a gap we note that in
the limit of large $L_0$ the sum can be approximated by the groundstate
potential term $n = 0$.

We may however analyze the same correlation also in terms of the transfer
matrix in one of the two spatial directions, say $\mu = 1$. Projecting to zero
momentum in the $\mu = 2$ direction we arrive at
\begin{equation}
  \sum_{x_2 = 0}^{L - 1} G \left( x_1, x_2 \right) = \sum_{n \geqslant 0} |v_n
  \left|^2 \mathe^{- \tilde{E}_n x_1} . \right.
\end{equation}
Here $v_n$ are non-trivial matrix elements and we have assumed $x_1 \ll L$,
$\tilde{E}_n L \gg 1$ to have purely exponential decay. The energy levels
$\tilde{E}_n$ refer to flux states created by the Polyakov line operator and
are thus expected to be asymptotically proportional to its length $L_0$.

Moreover, in {\cite{Luscher:2004ib}} L\"uscher and Weisz have shown that the
string state energies contain enough information to reconstruct the complete
correlator
\begin{equation}
  G \left( \vec{x} \right) = \sum_{n \geqslant 0} |v_n |^2 2 r \left(
  \frac{\tilde{E}_n}{2 \pi r} \right) K_0 \left( \tilde{E}_n r
  \right) \hspace{1em} \left( r = | \vec{x} | \right) .
\end{equation}
As one may expect form the presence of $r$ this formula holds in the continuum
theory and $K_0$ denotes a Bessel function. As $K_0$ decays exponentially for
large arguments, in particular the ground state energy $\tilde{E}_0$ controls
the large distance fall-off of $G \left( \vec{x} \right)$ with an asymptotic
area law.

\section{Effective string theory}

In effective string theories the fundamental degrees of freedom consist of a
two dimensional surface -- the time evolution of the string -- embedded in $D
\geqslant 3$ dimensional space. It furnishes an approximate description of
large Wilson loops (or Polyakov loop pairs) in gauge theories with the loop
position entering as the locus of Dirichlet boundary conditions for the edge
of the surface. The functional integral over all surface configurations with a
general action respecting the usual symmetries leads to a non-renormalizable
theory {\cite{Luscher:2004ib}} that is in some sense similar to chiral
perturbation theory. It predicts for instance an expansion of the lowest
potentials $V_n$ in (\ref{pots}) in powers of $r^{- 1}$. In each order only a
finite number of free parameters of the general action enter. In the closed
string interpretation expansions of the lowest $\tilde{E}_n$ in powers of
$L_0^{- 1}$ can be derived.

It turns out that for $D = 3$ the following result holds for the ground state
energy,
\begin{equation}
  \tilde{E}_0 = \sigma L_0 - \frac{\pi}{6 L_0} - \frac{\pi^2}{72 \sigma L_0^3}
  - \frac{\pi^3}{432 \sigma^2 L_0^5} + \Omicron \left( L_0^{- 7} \right) .
  \label{E0exp}
\end{equation}
Note that to this order the string tension $\sigma$ is the only dimensionful
parameter that enters. The $L_0^{- 5}$ has been shown to be parameter free
(`universal') (for $D = 3$) in {\cite{Aharony:2009gg}} while for the earlier
terms this result was already given in {\cite{Luscher:2004ib}}. Note that the
$L_0^{- 1}$ term (one loop in string theory) is the so-called L\"uscher term
that has been known for a long time {\cite{Luscher:1980fr}},
{\cite{Luscher:1980ac}}. We note that also the absence of even inverse powers
seems to be a nontrivial result.

In {\cite{Arvis:1983fp}} the Nambu-Goto geometric string action is used
instead of a general effective action restricted by symmetries only. This
leads to the formal all-order result
\begin{equation}
  z^2 = s^2 \left( 1 - \frac{1}{3 s} \right), \hspace{1em} s = \frac{\sigma
  L_0^2}{\pi}, \hspace{1em} z = \frac{\tilde{E}_0 L_0}{\pi}, \hspace{1em} s, z
  \rightarrow \infty \label{NG}
\end{equation}
which upon expansion implies the terms exhibited explicitly in (\ref{E0exp}).
Due to inconsistencies in the Nambu-Goto theory, the higher orders are not
expected to be correct, but at some point free coefficients that can
distinguish between different gauge models are expected to appear.

\section{Fitting string theory to new data}

In {\cite{Korzec:2012fa}} we have given results for $\tilde{E}_0$ for Z(2)
gauge theory at one lattice spacing with the string tension in lattice units
$\sigma \cong 0.0044$, $L = 64$, and $L_0 = 6 \ldots 28$. To have better
control over the continuum limit we here report on additional data for an
approximately two times smaller lattice spacing at $\beta = 0.75146$. They are
collected in table \ref{tab}. An iteration of the surface algorithm applied is
detailed in {\cite{Korzec:2012fa}} and corresponds to an order of Volume CPU
effort comparable to a sweep of a local algorithm. Again, as for the first
series of measurements, we found long plateaus for effective masses
$\tilde{E}_0$ and the extraction of numerical values proceeded as before.

\begin{table}
\begin{center}
  \begin{tabular}{|l|l|l|l|l|l|}
    \hline
    $L_0$ & $\tilde{E}_0$ & stat & $L_0$ & $\tilde{E}_0$ & stat\\
    \hline
    12 & 0.080998(6) & 85 & 36 & 0.381209(55) & 69\\
    \hline
    16 & 0.139831(9)$^{}$ & 74 & 40 & 0.426695(82) & 58\\
    \hline
    20 & 0.192035(20) & 33 & 44 & 0.47173(11) & 57\\
    \hline
    24 & 0.241174(30) & 33 & 48 & 0.51682(14) & 54\\
    \hline
    28 & 0.288694(43) & 32 & 52 & 0.56187(18) & 52\\
    \hline
    32 & 0.335173(39) & 72 & 56 & 0.60661(24) & 48\\
    \hline
  \end{tabular}
\end{center}
  \caption{New data complementing those listed in \cite{Korzec:2012fa}.
  Parameters are $\beta = 0.75146$, $L = 128$ and the statistics is given as
  multiples of $10^6$ iterations.}
  \label{tab}
\end{table}

\begin{figure}
\begin{center}
  \includegraphics[width=.8\textwidth]{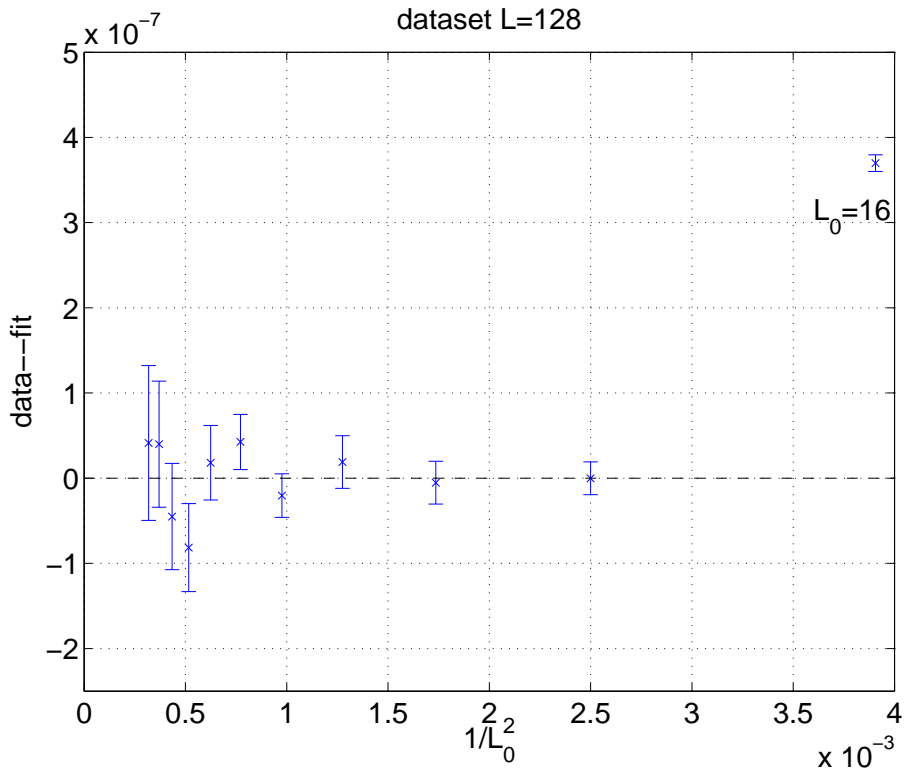}
  \caption{Difference between data and fit (\protect\ref{NGfit}).}
  \label{fig}
\end{center}
\end{figure}

The Nambu Goto result (\ref{NG}) would imply that the data can be fitted by
only two terms
\begin{equation}
  \frac{\tilde{E}_0^2}{L_0^2} = \sigma^2 + c_1 \frac{1}{L_0^2}, \label{NGfit}
\end{equation}
where in addition it even fixes $c_1$ by
\begin{equation}
  r = \left( \frac{3 c_1}{\pi \sigma} \right)^2 = 1. \label{rdef}
\end{equation}
We perform such a fit with a free value $c_1$ and, to have visible error bars
at all, plot immediately the deviation from the fit in figure \ref{fig}. In
table \ref{tab2} we give the achieved $\chi^2$ values depending on how small
$L_0$ we include in the fit. We see completely acceptable and stable fits under
varying $L_{0, \min} = 20, 24$, but find that there is no way to include $L_0 =
16$. The resulting values of $r$ are compatible with one, and hence at this
lattice spacing the Nambu Goto form (\ref{NG}) is consistent with our data for
$L_0 \geqslant 20 a$. This is in contrast to the lattices in
{\cite{Korzec:2012fa}} which are physically matched with just a
coarser lattice spacing. There $r$ had a significant 1\% deviation from one, whose
interpretation as a lattice artefact is hence confirmed.

\begin{table}
\begin{center}
  \begin{tabular}{|l|l|l|l|}
    \hline
    $L_{0, \min}$ & $\chi^2 / \tmop{dgf}$ & $\sigma$ & $r$\\
    \hline
    24 & 6.5/7 & 0.0109986(15) & 0.9986(41)\\
    \hline
    20 & 6.5/8 & 0.0109986(10) & 0.9987(20)\\
    \hline
    16 & 110/9 &  & \\
    \hline
  \end{tabular}
\end{center}
  \caption{Resulting fit parameters (\protect\ref{NGfit}), (\protect\ref{rdef}) when including
  $L_0 \geqslant L_{0, \min}$.}
  \label{tab2}
\end{table}

\section{Conclusions}

We have used our all-oder strong coupling, or random surface simulation
method, to investigate the energy of string states at various length that is
closely related to the large distance behavior of the potential in three
dimensional Z(2) gauge theory. The data fit well with the prediction of the
Nambu Goto effective string theory. It turns out, however, that our fit is
sensitive to only those terms in the prediction that are universal to all
effective string actions with the right symmetries. This is seen if we compare
our typical error $\delta \tilde{E}_0 \approx 2 \times 10^{- 5}$ with the size
of the first (presumably) non-universal term implied by (\ref{NG}) that reads
$5 \pi^4 / \left( 10368 \sigma^3 L_0^7 \right) \approx 10^{- 6}$, both for
$L_0 = 20$. It is the previous term that is borderline relevant by making a
contribution of about the same size as $\delta \tilde{E}_0$. The $L_0^{- 3}$
term is significantly tested by our result which also clearly and nicely
supports the general behavior of the string theory asymptotics as well as the
very abrupt break-away of the data from this form for smaller separations.

\end{document}